\documentclass[conference]{IEEEtran}

\pdfoutput=1

\newlength{\figwidth}
\usepackage{color}
\usepackage[dvipsnames]{xcolor}
\usepackage{pgfpages}  
\usepackage{pgfplots}
\usepackage{grffile}
\usepackage{tikz}
\usetikzlibrary{arrows}
\pgfplotsset{compat=newest}
\usetikzlibrary{plotmarks,patterns,shapes,fit}

\usepackage[nosort]{cite}        
\usepackage{url} 
\usepackage{amsmath}
\usepackage{bbm}
\usepackage{graphicx}
\usepackage{paralist}
\usepackage{fancyref}
\usepackage{vmr-symbols-vecbold}
\usepackage{standard-macros}
\usepackage{mathbbol}

\DeclareSymbolFontAlphabet{\amsmathbb}{AMSb}%

\newcommand{\lro}[1]{\lefto({#1}\right)}																
\newcommand{\lrbo}[1]{\lefto \lbrace {#1} \right \rbrace}															

\newcommand{\lr}[1]{\left({#1}\right)}																
\newcommand{\lrb}[1]{\left \lbrace {#1} \right \rbrace}															
\newcommand{\lrh}[1]{\left [ {#1} \right ]}																				

\safemath{\dopplerspread}{B_D}																								
\safemath{\delayspread}{T_D}																									
\safemath{\nc}{n\sub{c}}																										
\safemath{\nd}{n\sub{d}}																										
\safemath{\ntx}{n\sub{t}} 																											
\safemath{\nrx}{n\sub{r}}																											
\safemath{\ntxt}{\tilde{n\sub{t}}}																											
\safemath{\cb}{\ensuremath{L}} 																								
\safemath{\cl}{\ensuremath{n}} 																								
\safemath{\txanto}{{\ensuremath{\tilde{m}_t}}} 																		
\safemath{\cs}{M} 																														
\safemath{\idPustm}{\ensuremath{S_{k}}}
\safemath{\error}{\ensuremath{\epsilon}} 																				
\safemath{\eexp}{\ensuremath{\mathcal{E}}} 																			
\safemath{\nsubc}{n\sub{s}}			 																						
\safemath{\nofdm}{n\sub{o}} 																									
\safemath{\bc}{\ensuremath{B_c}} 																							
\safemath{\ts}{\ensuremath{T_s}} 																							
\safemath{\nrb}{\ensuremath{n_{rb}}} 																						
\safemath{\nres}{\ell}
\newcommand{\cgauss}[2]{\mathcal{CN}\lro{\ensuremath{#1, #2}  }}   								
\safemath{\maxk}{M^*\lr{\nres, \nsubc, \nofdm, \epsilon, \rho}}
\safemath{\Rmax}{R^*}
\safemath{\Emin}{E\sub{b}^*/N_0}
\safemath{\np}{\ensuremath{n\sub{p}}}
\safemath{\code}{\ensuremath{\mathcal{C}}}
\safemath{\err}{\ensuremath{\epsilon}}
\safemath{\rp}{\ensuremath{\rho\sub{p}}}
\safemath{\rd}{\ensuremath{\rho\sub{d}}}
\safemath{\Mr}{\ensuremath{M\sub{r}}}
\safemath{\hh}{\ensuremath{\hat{\randvech}}}
\safemath{\hhr}{\ensuremath{\hat{\vech}}}
\safemath{\mI}{\ensuremath{i\lro{\randvecy ; \randvecx}}} 				

\safemath{\randveca}{\bm{A}}
\safemath{\randvecb}{\bm{B}}
\safemath{\randvecc}{\bm{C}}
\safemath{\randvecd}{\bm{D}}
\safemath{\randvece}{\bm{E}}
\safemath{\randvecf}{\bm{F}}
\safemath{\randvecg}{\bm{G}}
\safemath{\randvech}{\bm{H}}
\safemath{\randveci}{\bm{I}}
\safemath{\randvecj}{\bm{J}}
\safemath{\randveck}{\bm{K}}
\safemath{\randvecl}{\bm{L}}
\safemath{\randvecm}{\bm{M}}
\safemath{\randvecn}{\bm{N}}
\safemath{\randveco}{\bm{O}}
\safemath{\randvecp}{\bm{P}}
\safemath{\randvecq}{\bm{Q}}
\safemath{\randvecr}{\bm{R}}
\safemath{\randvecs}{\bm{S}}
\safemath{\randvect}{\bm{T}}
\safemath{\randvecu}{\bm{U}}
\safemath{\randvecv}{\bm{V}}
\safemath{\randvecw}{\bm{W}}
\safemath{\randvecx}{\bm{X}}
\safemath{\randvecy}{\bm{Y}}
\safemath{\randvecz}{\bm{Z}}
\safemath{\randvecphi}{\bm{\Phi}}

\safemath{\randmatA}{\amsmathbb{A}}
\safemath{\randmatB}{\amsmathbb{B}}
\safemath{\randmatC}{\amsmathbb{C}}
\safemath{\randmatD}{\amsmathbb{D}}
\safemath{\randmatE}{\amsmathbb{E}}
\safemath{\randmatF}{\amsmathbb{F}}
\safemath{\randmatG}{\amsmathbb{G}}
\safemath{\randmatH}{\amsmathbb{H}}
\safemath{\randmatI}{\amsmathbb{I}}
\safemath{\randmatJ}{\amsmathbb{J}}
\safemath{\randmatK}{\amsmathbb{K}}
\safemath{\randmatL}{\amsmathbb{L}}
\safemath{\randmatM}{\amsmathbb{M}}
\safemath{\randmatN}{\amsmathbb{N}}
\safemath{\randmatO}{\amsmathbb{O}}
\safemath{\randmatP}{\amsmathbb{P}}
\safemath{\randmatQ}{\amsmathbb{Q}}
\safemath{\randmatR}{\amsmathbb{R}}
\safemath{\randmatS}{\amsmathbb{S}}
\safemath{\randmatT}{\amsmathbb{T}}
\safemath{\randmatU}{\amsmathbb{U}}
\safemath{\randmatV}{\amsmathbb{V}}
\safemath{\randmatW}{\amsmathbb{W}}
\safemath{\randmatX}{\amsmathbb{X}}
\safemath{\randmatY}{\amsmathbb{Y}}
\safemath{\randmatZ}{\amsmathbb{Z}}
\safemath{\randmatSigma}{\mathbb{\Sigma}}
\safemath{\randmatPhi}{\mathbb{\Phi}}
\safemath{\randmatLambda}{\mathbb{\Lambda}}

\safemath{\matSigma}{\bm{\Sigma}}
\safemath{\matPhi}{\bm{\Phi}}
\safemath{\matLambda}{\bm{\Lambda}}

\def\bie{\begin{IEEEeqnarray}{rCl}}
\def\eie{\end{IEEEeqnarray}}

\usepackage{header}

\usepackage{kolor}

\usepackage{kontext}

\usepackage{konfig}

\let\marginnote\undefined

\newcommand{\marginnote}[1]{\marginpar{\framebox{\framebox{#1}}}}

\def\vect#1{\textup{vec}(#1)}
\let\geq\geqslant
\let\leq\leqslant 
\usepackage[scaled=0.9]{helvet}

\usepackage{tabularx,booktabs}
\newcolumntype{L}[1]{>{\raggedright\let\newline\\\arraybackslash\hspace{0pt}}m{#1}}
\newcolumntype{C}[1]{>{\centering\let\newline\\\arraybackslash\hspace{0pt}}m{#1}}
\newcolumntype{R}[1]{>{\raggedleft\let\newline\\\arraybackslash\hspace{0pt}}m{#1}}
\usepackage{boldline}

\safemath{\txant}{M\sub{t}}
\safemath{\rxant}{M\sub{r}}
\safemath{\cohtime}{n\sub{c}}
\safemath{\npilot}{n\sub{p}}
\safemath{\ndata}{n\sub{d}}
\let\snr\undefined
\safemath{\snr}{\rho}
\safemath{\matdata}{\rmatX_{\ell}^{(\text{d})}}
\safemath{\matpilot}{\rmatX_{\ell}^{(\text{p})}}
\safemath{\matdatabar}{\bar{\rmatX}_{\ell}^{(\text{d})}}
\safemath{\matpilotbar}{\bar{\rmatX}_{\ell}^{(\text{p})}}

\safemath{\matdatarx}{\rmatY_{\ell}^{(\text{d})}}
\safemath{\matpilotrx}{\rmatY_{\ell}^{(\text{p})}}

\safemath{\vecdata}{\vecx_{\ell,k}^{(\text{d})}}
\safemath{\vecpilot}{\vecx_{\ell,k}^{(\text{p})}}
\safemath{\vecdatarx}{\vecy_{\ell,k}^{(\text{d})}}
\safemath{\vecpilotrx}{\vecy_{\ell,k}^{(\text{p})}}

\begin{document}

\IEEEoverridecommandlockouts

\title{Pilot-Assisted Short-Packet Transmission over Multiantenna Fading Channels: A 5G Case Study}
  \author{\IEEEauthorblockN{Guido Carlo Ferrante$^{1}$, Johan \"Ostman$^{1}$, Giuseppe Durisi$^{1}$, and Kittipong Kittichokechai$^{2}$}\\[-3mm]
\IEEEauthorblockA{
$^1$Chalmers University of Technology, Gothenburg, Sweden\\$^2$Ericsson Research, Ericsson AB,  Sweden }
%
\thanks{This work was partly supported by the Swedish Research Council under grants 2014-6066 and 2016-03293.
The simulations were performed in part on resources provided by the Swedish National Infrastructure for Computing (SNIC) at C3SE.}
}
%
\maketitle
\begin{abstract}\boldmath
  Leveraging recent results in finite-blocklength information theory, we investigate  the problem of designing a control channel in a 5G system.
 The setup involves the transmission, under stringent latency and reliability constraints, of a short data packet containing a small information payload, over a propagation channel that offers limited frequency diversity and no time diversity.
 We present an achievability bound, built upon the random-coding union bound with parameter $s$ (\emph{Martinez \& {Guill\'en i F\`abregas}, 2011}), which relies on quadrature phase-shift keying  modulation, pilot-assisted transmission to estimate the fading channel, and scaled nearest-neighbor decoding at the receiver.
Using our achievability bound, we determine how many pilot symbols should be transmitted to  optimally trade  between channel-estimation errors and rate loss due to pilot overhead. 
Our analysis also reveals the importance of using multiple antennas at the transmitter and/or the receiver to provide the spatial diversity needed to meet the stringent reliability constraint.
\end{abstract}

\section{Introduction}
\label{sec:introduction}
Ultra-reliable low-latency communication (URLLC) is one of the new use cases that will be supported in 5G~\cite{itu-10-17}. 
It involves the transmission of short packets, under latency and reliability constraints that are much more stringent than the ones satisfied by traditional mobile broadband applications.
Possible applications include factory automation and traffic safety.

Classical information-theoretic metrics, such as the ergodic and the outage capacity, are not suitable to design URLLC links, because they rely on the assumption of large blocklength, which is typically not compatible with the latency requirements in URLLC links~\cite{durisi16-09a}.
Instead, the problem of optimally designing such systems can be tackled in a fundamental fashion using the finite-blocklength information-theoretic tools developed by Polyanskiy \emph{et al.}~\cite{polyanskiy10-05a}.

These tools have recently enabled the characterization of the maximum coding rate achievable, for a given blocklength and a given error probability, over quasi-static fading channels~\cite{yang14-07c}, and over multiple-input multiple-output (MIMO) Rayleigh block-fading channels~\cite{durisi16-02a}. 
They have also been used to determine optimum power-control strategies in the presence of channel-state information (CSI) at the transmitter~\cite{yang15-09a}, and to bound the rates achievable with pilot-assisted transmission (PAT) followed by scaled nearest-neighbor (SNN) decoding at the receiver for the single-input single-output (SISO) Rician block-fading channel~\cite{ostman17-12a}. 
In~\cite{ostman17-12a}, the design of actual channel coding schemes approaching the bounds is also discussed.

\paragraph*{Contributions} 
In this paper, we generalize the analysis in~\cite{ostman17-12a} to the case of multiple-antenna transmissions. 
Specifically, we present an upper bound on the packet error probability attainable at a given blocklength using a channel code of a fixed rate, when communicating over a MIMO block-fading channel. 

As in~\cite{ostman17-12a}, we assume PAT and SNN decoding at the receiver. 
However, differently from~\cite{ostman17-12a}, where the analysis relies on the transmission of spherical codes, we focus in this paper on the rates achievable using quadrature phase-shift keying (QPSK) modulation, which is more practically relevant, and also a natural choice given the low levels of spectral efficiency at which 5G URLLC links are expected to operate.
We also consider the use of an Alamouti inner code~\cite{alamouti98-10a} at the transmitter, which 
constrains the transmit antennas to provide only spatial diversity, which may be crucial to achieve high reliability levels.

Our bound is not in closed form; its evaluation require Monte Carlo simulations, which may be time consuming if the target error probability is low. 
To partially overcome this issue, we present an accurate saddlepoint approximation~\cite{jensen95,scarlett14-05a} of our bound, which, although not in closed form either, can be computed more efficiently that the  bound, because its complexity  does not increase with the number of diversity branches available in the channel.

Finally, we use our bound to shed lights on the optimal design of a control channel in a 5G  system, where the payload is assumed to be $30$ bits, the target packet error probability is $10^{-5}$ and the data packet consists of multiple resource blocks (RBs) in frequency, so as to minimize latency. 
Furthermore, the spacing between the RBs is chosen so as to optimally exploit the frequency diversity offered by the channel. 
The coherence time and the coherence bandwidth of the block-fading model are chosen so as to match the ones prescribed by the extended pedestrian type~A (EPA) 5\,Hz \cite{3GPP-TS-36.104} and the tapped delay line type-C (TDL-C) 300\,ns--3\,km/h \cite{3GPP-TR-38.901} channel models. 
Furthermore, the number and the distribution of the RBs in frequency as well as the number of pilot symbols are optimized. 
We analyze how the performance of a single-input multiple-output (SIMO) system depends on the number of available receive antennas.
We also illustrate that the sensitivity of the Alamouti scheme to imperfect channel estimation makes this scheme unsuitable for transmission over channels exhibiting a large amount of frequency selectivity.

\paragraph*{Notation}
We shall denote vectors and matrices by bold lower and uppercase letters, such as $\vecx$ and $\rmatX$, respectively. 
The identity matrix of size $a\times a$ is written as $\rmatI_{a}$.
The distribution of a circularly-symmetric complex Gaussian random variable with variance $\sigma^2$ is denoted by $\cgauss{0}{\sigma^2}$. 
The superscripts~ $\conj{\lro{\cdot}}$, $\tp{\lro{\cdot }}$, and $\herm{\lro{\cdot }}$ denote conjugation, transposition, and Hermitian transposition, respectively. We write $\log(\cdot)$ and $\log_2(\cdot)$ to denote the natural logarithm and the logarithm to the base $2$, respectively.
Finally, $\lrh{a}^+$ stands for $\max\lrbo{0, a}$, $Q\lro{\cdot}$ denotes the Gaussian $Q$-function, $\vecnorm{\cdot}$ the $\ell^2$-norm, $\frobnorm{\cdot}$ the Frobenius norm, and $\Ex{}{\cdot}$ the expectation operator.


\section{System Model}
\label{sec:sysmod}
\subsection{Input-Output Relation} 
\label{sec:input_output_relation_and_definition_of_a_code}

We consider a discrete-time MIMO  block-fading channel with $M\sub{t}$ transmit and $M\sub{r}$ receive antennas. 
Let $n\sub{c}$ be the size of each coherence block, i.e., the number of channel uses over which the channel stays constant.
We assume that each codeword of length $n$ spans $L$ coherence blocks, i.e., $n=Ln\sub{c}$. 
We shall refer to $L$ as the number of diversity branches.
The signal received during block $\ell$ is 
\begin{equation}\label{eq:io}
\bm{Y}_{\ell} = \bm{H}_{\ell} \bm{X}_{\ell} + \bm{W}_{\ell},
\end{equation}
where $\bm{Y}_{\ell}\in\complexset^{M\sub{r} \times n\sub{c}}$ is the channel output,
$\bm{H}_{\ell}\in\complexset^{M\sub{r} \times M\sub{t}}$ is the matrix containing the fading coefficients in the $\ell$th coherence block, $\bm{X}_{\ell}\in\complexset^{M\sub{t} \times n\sub{c}}$ is the channel input, and $\bm{W}_{\ell}\in\complexset^{M\sub{r} \times n\sub{c}}$ is the AWGN matrix. 
The noise matrices $\{\bm{W}_{\ell}\}$ have independent and identically distributed (\iid) entries drawn from $\jpg(0,1)$ and are independent across $\ell$.
The fading matrices $\{\rmatH_{\ell}\}$ are also \iid over $\ell$; their distribution is, however, arbitrary.
Furthermore, we assume that  $\{\rmatH_{\ell}\}$ and $\{\rmatW_{\ell}\}$ are independent, and that they do not depend on $\{\bm{X}_\ell\}$.
We next define the notion of a channel code.

\begin{dfn} An $(n,M,\epsilon)$-code consists of:
  \begin{itemize}
    \item An encoder $f:\{1,\dots,M\}\rightarrow \complexset^{M\sub{t}\times n}$ that maps the message $J$, which is uniformly distributed on the set $\{1,\dots,M\}$ to a codeword $\rmatC_m=f(J) \in \complexset^{\txant\times n}$ in the codebook set $\{\rmatC_1,\dots, \rmatC_M\}$.
    Each codeword satisfies the power constraint $\frobnorm{\rmatC_m}^2\leq \cohtime\snr$, $m=1,\dots, M$.
    \item A decoder $g: \complexset^{\rxant\times n}\rightarrow \{1,\dots,M\} $ that maps the channel output $\rmatY=[\rmatY_1,\dots \rmatY_{L}]$ to a message estimate $\widehat{J}=g(\rmatY)$.
    The decoder satisfies the average packet error probability constraint
    \begin{equation}\label{eq:average_prob}
     \Pr\{\widehat{J}\neq J\}\leq \epsilon.
    \end{equation}
  \end{itemize}
\end{dfn}

The maximum coding rate $R^*(n,\epsilon)$ for a given blocklength $n$ and a given error probability $\epsilon$ is the largest rate achievable using $(n,M,\epsilon)$-codes:
\begin{equation}\label{eq:max_coding_rate}
  R^*(n,\epsilon)=\sup\lefto\{\frac{\log_2 M}{n}\sothat\exists (n,M,\epsilon)\text{-code}\right\}.
\end{equation}
Similarly, we define the minimum error probability $\epsilon^*(n,R)$ achievable using codes of blocklength $n$ and rate $R=n^{-1}\log_2(M)$ as
\begin{equation}\label{eq:min_err_prob}
  \epsilon^{*}(n,R)=\inf\lefto\{\epsilon\sothat \exists (n,\ceil{2^{nR}},\epsilon)
  \text{-code}\right\}.
\end{equation}
This quantity is often studied as a function of the energy per bit  normalized by
the noise spectral density, $E\sub{b}/N_0=\snr/R$.

\subsection{PAT and SNN Decoding} 
\label{sec:pat_and_snn_decoding}
We assume that each input matrix $\rmatX_\ell$ is of the form $\rmatX_\ell=[\matpilot \matdata]$ where $\matpilot\in\complexset^{\txant\times\npilot}$. 
Here, $\matpilot$, with $\txant\leq \npilot<\cohtime$, is a deterministic matrix containing orthogonal pilot sequences in each row.
Specifically, we assume that $\matpilot\herm{\bigl(\matpilot\bigr)}
=(\snr\npilot/\txant)\rmatI_{\txant}$.
The matrix $\matdata\in\complexset^{\txant\times\ndata}$, where $\ndata=\cohtime-\npilot$, contains the data symbols.

Let $\matpilotrx$ and $\matdatarx$ be the matrices containing the received samples that correspond to the pilot and the data symbols within the $\ell$th coherence block, respectively.
Given $\matpilotrx$ and $\matdatarx$, the receiver computes the maximum likelihood (ML) estimate $\widehat{\rmatH}_{\ell}$ of the fading matrix $\rmatH_{\ell}$ as
\begin{equation}\label{eq:ml_channel_estimation}
  \widehat{\rmatH}_{\ell}=\frac{\txant}{\snr\npilot}\matpilotrx\herm{\bigl(\matpilot\bigr)}.
\end{equation}
Then, the decoder produces as output the message 
\begin{equation}\label{eq:mismatched_decoding_rule}
  \widehat{J}=\argmax_{1 \leq m\leq M} q^{(L)}(\rmatC_{m},\rmatY)
\end{equation}
where 
\begin{equation}\label{eq:total_decoding_metric}
  q^{(L)}(\rmatX,\rmatY)=\prod_{\ell=1}^{L}q(\rmatX_\ell,\rmatY_{\ell})
\end{equation}
with $\rmatX=[\rmatX_1,\dots,\rmatX_{L}]$ and
\begin{equation}\label{eq:blockwise_decoding_metric}
  q(\rmatX_{\ell},\rmatY_{\ell})=\prod_{k=1}^{\ndata} \exp\bigl(-\vecnorm{\vecdatarx-\widehat{\rmatH}_\ell\vecdata}^2\bigr)
\end{equation}
is the SNN decoding metric.
Here, $\vecdatarx$ and $\vecdata$ denote the $k$th column of the matrices $\matdatarx$ and $\matdata$, respectively.

Some remarks on~\eqref{eq:blockwise_decoding_metric} are in order. 
When $\widehat{\rmatH}_{\ell}=\rmatH_{\ell}$, i.e., when perfect CSI is available at the receiver, the SNN decoding metric $q(\rmatX_{\ell},\rmatY_{\ell})$ in~\eqref{eq:blockwise_decoding_metric} is equivalent to the ML metric, which is optimal in the sense that it minimizes the error probability $\Pr\{\widehat{J}\neq J\}$. 
However, using this rule for the case of inaccurate CSI considered in this paper yields a mismatch. 

The transceiver architecture just described, which relies on PAT, on ML channel estimation, and on SNN decoding, and which we shall refer to as PAT-ML-SNN coding scheme, is ubiquitous in current wireless systems, although suboptimal.
Treating the channel estimate as perfect enables the use of the ``coherent'' decoding rule~\eqref{eq:mismatched_decoding_rule}--\eqref{eq:blockwise_decoding_metric}, 
whose performance can be approached in practice using good channel codes for the AWGN channel.

\section{Bounds on the Error Probability} 
\label{sec:bounds_on_the_error_probability}
The performance of the PAT-ML-SNN coding scheme just introduced can be analyzed using the mismatch-decoding framework~\cite{merhav94-11a}.
Specifically, our analysis is based on the RCUs achievability bound~\cite[Thm.~1]{martinez11-02a}, a relaxation of the RCU bound~\cite[Thm.~16]{polyanskiy10-05a} that recovers the generalized random-coding error exponent for mismatch detection introduced in~\cite{kaplan93-a}.
Our main result is given in the following theorem.

\begin{thm}\label{thm:RCUs_general}
Fix  an integer $1\leq\ndata<\cohtime$, a real number $s\geq 0$, and  a probability distribution $P_{\rmatX^{(\text{d})}}$ on $\complexset^{\txant \times \ndata}$ for which $\frobnorm{\rmatX^{(\text{d})}}^2\leq \snr\ndata$ \wpone when $\rmatX^{(\text{d})}\distas P_{\rmatX^{(\text{d})}}$.
Let the generalized information density $\imath_{s}(\rmatX_{\ell},\rmatY_{\ell})$ be defined as
\begin{equation} 
\imath_{s}(\bm{X}_{\ell},\bm{Y}_{\ell}) = \log\frac{q(\bm{X}_{\ell},\bm{Y}_{\ell})^{s}}{\mathbb{E}_{\bm{\bar{X}}_{\ell}}[q(\bm{\bar{X}}_{\ell},\bm{Y}_{\ell})^{s}]}
\end{equation}
where $\bar{\rmatX}_{\ell}=[\matpilotbar,\matdatabar]$ with $\matpilotbar$ an arbitrary pilot matrix satisfying the properties listed in Section~\ref{sec:pat_and_snn_decoding}, and~$\matdatabar \distas P_{\rmatX^{(\text{d})}}$.
The average error probability $\epsilon(n,R)$ achievable with the PAT-ML-SSN coding scheme described in Section~\ref{sec:pat_and_snn_decoding} is upper-bounded by
\begin{multline}\label{eq:RCUs_error}
  \epsilon(n,R)\leq \epsilon\sub{RCUs}(n,R)\\
  =\Ex{}{\exp\biggl(-\biggl[\biggl(\sum_{\ell=1}^L \imath_{s}(\bm{X}_\ell,\bm{Y}_\ell)\biggr)-\log(2^{nR}-1) \biggr]^{+} \biggr)}
\end{multline}
where $\rmatX_{\ell}$ is distributed as $\bar{\rmatX}_{\ell}$ and $\rmatY_{\ell}$ is the induced channel output according to~\eqref{eq:io}.
\end{thm}
\begin{IEEEproof}
  We consider all codebooks whose codewords have data symbols that are generated independently according to the product distribution built upon $P_{\rmatX^{(\text{d})}}$, and have the pilot symbols $\matpilot$ in each coherence interval.
  It follows from~\cite[Thm.~1]{martinez11-02a} that the error probability, averaged over all codebooks, achievable with the decoding rule~\eqref{eq:mismatched_decoding_rule} is upper-bounded by $\epsilon\sub{RCUs}(n,R)$ given in~\eqref{eq:RCUs_error}.
\end{IEEEproof}
\subsection{Saddlepoint Approximation} 
\label{sec:saddle_point_approximation}
When  $P_{\rmatX^{(\text{d})}}$ is taken as product distribution, i.e., $P_{\rmatX^{(\text{d})}}(\rmatX^{(\text{d})})=\prod_{k=1}^{\ndata}P_{\vecx}(\vecx^{(\text{d})}_k)$,
where $\vecx_k^{(\text{d})}$ stands for the $k$th column of $\rmatX^{(\text{d})}$---a choice we will focus on in the numerical results reported in Section~\ref{sec:numerical_results}---the generalized information density takes the following form:
\begin{IEEEeqnarray}{rCL}\label{eq:generalized_inf_density_iid}
  \imath_s(\rmatX_{\ell},\rmatY_{\ell})&=&\sum_{k=1}^{\ndata} \biggl\{
  -s \vecnorm{\vecdatarx-\widehat{\rmatH}_{\ell}\vecdata}^2\notag\\
  &&-\log \Ex{}{\exp\lefto(-s\vecnorm{\vecdatarx-\widehat{\rmatH}_{\ell}\vecdata}^2\right)}.\IEEEeqnarraynumspace
\end{IEEEeqnarray}
Even in this case, though, $\epsilon\sub{RCUs}(n,R)$, does not admit in general a closed-form expression.
This makes its computation challenging for low error probabilities.
To partly overcome this issue, we present next a saddlepoint approximation of~$\epsilon\sub{RCUs}(n,R)$, which we will show in Section~\ref{sec:numerical_results} to be remarkably accurate over a large range of channel and system parameters.
We obtain this approximation  by proceeding as in~\cite[Sec. IV.B]{martinez11-02a} (see also~\cite[Sec.~V]{scarlett14-05a}, where the error in the approximation is analyzed).
The resulting saddlepoint approximation of $\epsilon\sub{RCUs}(n,R)$ is
\begin{IEEEeqnarray}{rCL}
&&\epsilon\sub{RCUs}(n,R)
\approx \exp\Bigl({-L \bigl[E_0\lro{\hat{\tau},s}-\hat{\tau} E'_0\lro{\hat{\tau},s}}\bigr]\Bigr) \notag \\
&&\hspace{+15ex}\cdot \Big\{Q\Big({\hat{\tau}\sqrt{-LE''_0\lro{\hat{\tau},s}}}\Big)e^{-\frac{L}{2}E''_0\lro{\hat{\tau},s}\hat{\tau}^2}  \notag \\
&&\hspace{+5ex}+Q\Big({\lro{1-\hat{\tau}}\sqrt{-LE''_0\lro{\hat{\tau},s}}}\Big)e^{-\frac{L}{2}E''_0\lro{\hat{\tau},s}\lro{1-\hat{\tau}}^2} \Big\}\IEEEeqnarraynumspace   \label{eq:SP_RCUs}
\end{IEEEeqnarray}
where the Gallager's generalized $E_0$ function is
\begin{IEEEeqnarray}{rCl} \label{eq:E0}
E_0\lro{\tau,s} = -\log\Ex{}{e^{-\tau \imath_s(\rmatX_{1},\rmatY_{1})}},
\end{IEEEeqnarray}
and $E'_0$ and $E''_0$ denote the first and the second partial derivatives of $E_0$ with respect to $\tau$, respectively,
The parameter $\hat{\tau}$ in~\eqref{eq:SP_RCUs} is
\begin{IEEEeqnarray}{rCl} 
	\hat{\tau} = \argmax_{\tau \in \lro{0,1}} \lrbo{E_0\lro{\tau,s} - \tau \frac{\log\lro{2^{nR}-1}}{L}}.
\end{IEEEeqnarray}

A closed form expression for~\eqref{eq:E0} and its partial derivatives is in general not available.
Hence, we shall turn to Monte Carlo methods to evaluate~\eqref{eq:SP_RCUs}.
Note, that, due to the block-memoryless assumption, the numerical complexity of the approximation in~\eqref{eq:SP_RCUs} is independent of the number of diversity branches $L$.
In contrast, the complexity of the numerical evaluation of $\epsilon\sub{RCUs}(n,R)$ in~\eqref{eq:RCUs_error} increases with $L$.


\subsection{The SIMO Case} 
\label{sec:the_simo_case}
In the SIMO case, the SNN decoding metric~\eqref{eq:blockwise_decoding_metric} reduces to
\begin{equation}\label{eq:decoding_metric_simo}
  q(\vecx_{\ell},\rmatY_{\ell})=\prod_{k=1}^{\ndata} \exp\lefto(-\vecnorm{\vecdatarx-\widehat{\vech}_{\ell}x^{(\text{d})}_{\ell,k}}^2\right).
\end{equation}
To simplify the computation of~\eqref{eq:RCUs_error}, it is convenient to left-multiply the vector $\vecdatarx-\widehat{\vech}_{\ell}x_{\ell,k}$ by a unitary matrix whose first row is $\herm{\widehat{\vech}}_{\ell}/\vecnorm{\widehat{\vech}_{\ell}}$.
This corresponds to performing maximum-ratio combining at the receiver, based on the estimated CSI $\widehat{\vech}_{\ell}$.
Since all the entries of the rotated vectors but the first one do not depend on $x_{\ell,k}$ they can be dropped when solving~\eqref{eq:mismatched_decoding_rule}.
This is equivalent to applying the RCUs bound in Theorem~\ref{thm:RCUs_general} to the following setup: 
\begin{inparaenum}[i)]
  \item the channel~\eqref{eq:io} is replaced by the block-fading SISO channel 
  \begin{equation}\label{eq:equivalent simo} \tp{\vecy}_{\ell}=\frac{\herm{\widehat{\vech}_{\ell}}\vech_{\ell}}{\vecnorm{\widehat{\vech}_{\ell}}}\tp{\vecx}_\ell + \tp{\vecw}_{\ell}
  \end{equation}
where the vectors $\vecy_{\ell}$, $\vecx_{\ell}$, and $\vecw_{\ell}$ have $\ndata$ entries; furthermore, the vector ${\vech}_{\ell}$ contains the $\rxant$ channel coefficients in the $\ell$th coherence block and   $\widehat{\vech}_{\ell}$ is its ML estimate.
\item The decoding metric $q(\cdot,\cdot)$ in~\eqref{eq:blockwise_decoding_metric} is replaced by 
  \begin{equation}\label{eq:blockwise_decoding_metric_simo}
    q(\vecx_{\ell},\vecy_{\ell})=\prod_{k=1}^{\ndata} \exp\bigl(-\bigl\lvert y_{\ell,k}-\vecnorm{\widehat{\vech}_\ell}x_{\ell,k}\bigr\rvert^2\bigr).
  \end{equation}
  
\end{inparaenum}

\subsection{Spatial Diversity through Alamouti} 
\label{sec:spatial_diversity_through_alamouti}
We next focus on the $2\times 2$ MIMO setup, and discuss the scenario in which an Alamouti inner code is used to obtain spatial diversity from the two available transmit antennas.
Let $\vecx_{\ell} \in \complexset^{\ndata}$, with $\vecnorm{\vecx_{\ell}}^2\leq \ndata\snr/2$, be the vector of data symbols to be transmitted over the coherence block $\ell$.
Through this section, we shall assume that $\ndata$ is even.
The data matrix $\matdata \in \complexset^{2\times \ndata}$ is constructed as
\begin{equation}\label{eq:alamouti_input}
  \matdata = \mat \tp{\vecx}_{\ell} \\
      \tp{e(\vecx_{\ell})}
   \emat.
\end{equation}
Here, the function $e: \complexset^{\ndata}\rightarrow \complexset^{\ndata}$ maps an input vector $\veca$ into an output vector $\vecb$ according to the Alamouti rule~\cite{alamouti98-10a}
\begin{IEEEeqnarray}{rCL}
    &&[\vecb]_{2k-1} = [e(\veca)]_{2k-1}= \conj{[\veca]}_{2k} \\
    &&{[\vecb]}_{2k} =  [e(\veca)]_{2k}= -\conj{[\veca]}_{2k-1}
    \label{eq:alamouti_rule}
\end{IEEEeqnarray}
for $k=1,\dots, \ndata/2$.
Exploiting the structure of the data matrix, one can show that, for the case of perfect CSIR, the performance of this scheme is equal to that of a $1\times 4$ SIMO system where the power of the data symbols is halved.

This is, however, no longer the case when the CSI is acquired through pilot symbols, and hence, inaccurate.
In this case, the performance of this coding scheme can be analyzed by applying the RCUs bound in Theorem~\ref{thm:RCUs_general} to an equivalent channel and decoding metric we shall specify next.
Let $\tilde{\vecx}_{\ell,k}$ be a $2$-dimensional vector obtained from $\vecx_{\ell}$ by taking  the symbol in position $2k-1$ and the complex conjugate of the symbols in position $2k$, $k=1,\dots \ndata/2$.
We apply the RCUs bound to the equivalent channel
\begin{equation}\label{eq:equivalent_channel_alamouti}
  \vecy_{\ell,k}= \frac{1}{\frobnorm{\widehat{\rmatH}_{\ell}}} \left(\herm{\widehat{\rmatV}}_{\ell,1}\rmatV_{\ell,1} +\herm{\widehat{\rmatV}}_{\ell,2}\rmatV_{\ell,2}\right)\tilde{\vecx}_{\ell,k}+\vecw_{\ell,k}
\end{equation}
where 
\begin{equation}\label{eq:V_matrix}
  \rmatV_{\ell,j}=
  \mat
  h_{\ell,1,j} & h_{\ell,2,j} \\
  \conj{h_{\ell,2,j}} & -\conj{h}_{\ell,1,j}
  \emat
\end{equation}
with $h_{\ell,i,j}=[\rmatH_{\ell}]_{i,j}$, $i\in\{1,2\}$, $j\in\{1,2\}$.
The matrix $\widehat{\rmatV}_{\ell,j}$ is defined as in~\eqref{eq:V_matrix}, with the entries of $\rmatH_{\ell}$ replaced by the entries of its ML  estimate 
$\widehat{\rmatH_{\ell}}$.

Furthermore, we use the symbol-wise mismatch decoding metric
  \begin{equation}\label{eq:blockwise_decoding_metric_alamouti}
    q(\tilde{\vecx}_{\ell},\vecy_{\ell})=\prod_{k=1}^{\ndata} \exp\bigl(-\bigl\lvert y_{\ell,k}-\frobnorm{\widehat{\rmatH}_\ell}\,\tilde{x}_{\ell,k}\bigr\rvert^2\bigr).
  \end{equation}


\section{Numerical Results} 
\label{sec:numerical_results}

\begin{figure}\centering
\includegraphics{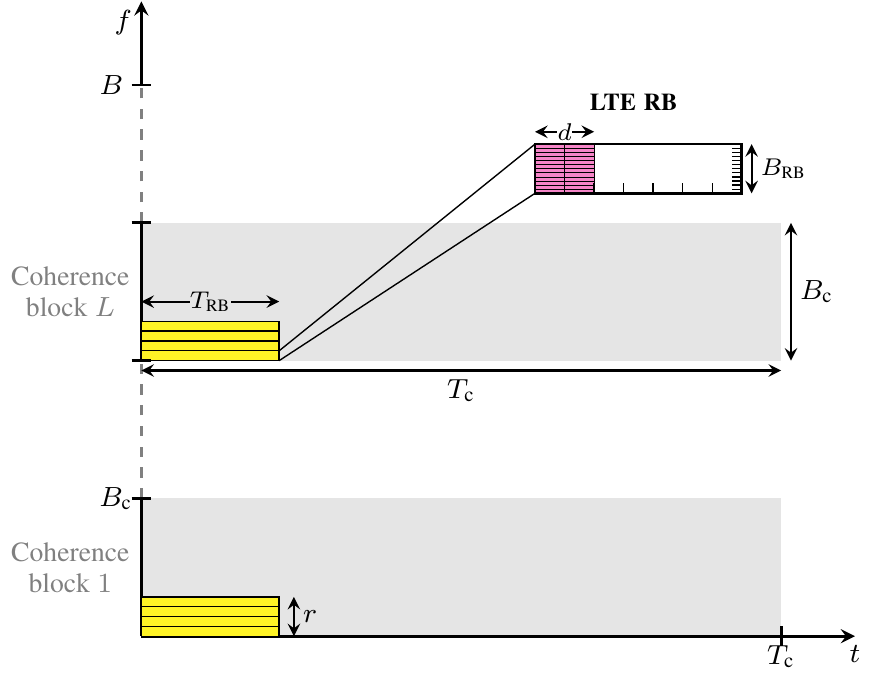}
\caption{Channel and signal properties in the time-frequency plane.}
\label{fig:channel-scheme}
\end{figure}

\begin{table}[t]\centering
\caption{Channel (upper half) and signal (bottom half) parameters.}
\label{tab:channel-parameters}
\begin{tabular}{C{0.95cm}C{2.85cm}C{1.5cm}C{1.75cm}}
\toprule
\bf Symbol 			& \bf Parameter		& \bf EPA 5\,Hz 	& \bf TDL-C 300\,ns--3\,km/h \\
\midrule
$B\sub{c}$			& 50\% coh. bandwidth 	& $4.4$ MHz 		& $0.66$ MHz \\
$T\sub{c}$			& 50\% coh. time 		& $85$ ms		& $85$ ms \\
$L\sub{max}$		& Max no. div. branches 	& $4$			& $30$ \\
\midrule
$L$				& No. div. branches 		& $4$			& $12$ \\
$n\sub{c}$			& Coh. block size		& $72$ 		& $24$ \\
$n$				& Blocklength		& $288$		& $288$ \\
\bottomrule
\end{tabular}
\end{table}

We consider a target packet error probability of $\epsilon=10^{-5}$, in line with the specifications for URLLC \cite{3GPP-TR-38.913},  a payload of $k=30$ bits, which models the so-called \textit{compact downlink control information} (DCI) \cite{R1-1720997-Ericsson}, and a blocklength $n=288$ symbols.
This results in a rate of $R=k/n=0.104$ bit/channel use.
Our goal is to identify the system parameters (number of antennas, number of pilot symbols, and distributions of the symbols in the time-frequency plane) that allow us to meet the above requirements.
To do so, we  use the RCUs bound in Theorem~\ref{thm:RCUs_general}.\footnote{Throughout this section, we set $s=1$ in~\eqref{eq:RCUs_error} for simplicity.
An optimization over $s\geq 0$ is left for future works.}

For the sake of concreteness, we take as input distribution independent QPSK signaling  and assume that the pilot and the data symbols are transmitted at the same power level.
Also we  focus on $1\times \rxant$ SIMO, with $\rxant \in \{1,2,4\}$ and on $2\times 2$ MIMO with Alamouti.


We assume the use of orthogonal frequency-division multiplexing (OFDM) with an LTE numerology.
This means that each codeword is assigned a number of RBs, each one consisting of $d$ OFDM symbols spanning $12$ subcarriers. 
For a typical downlink control channel transmission in LTE, $d$ is between $1$ and $3$.
As shown in Fig.~\ref{fig:channel-scheme}, we allow the RBs to be separated in frequency, but not in time, to benefit from frequency diversity and to limit the transmission delay.

We consider two channel models, which yield a different number of available diversity branches: the EPA 5\,Hz \cite{3GPP-TS-36.104} and the TDL-C 300\,ns--3\,km/h  \cite{3GPP-TR-38.901}.
To map these channel models into the block-memoryless fading model~\eqref{eq:io}, we compute their coherence bandwidth $B\sub{c}$ and their coherence time $T\sub{c}$.
These values are given in Table~\ref{tab:channel-parameters}.
Note that the system bandwidth in LTE is $B=20\MHz$ and that an RB lasts $0.5\ms$ and occupies $B\sub{RB}=180\kHz$.
This means that the two channels offer no time diversity and a maximum number of diversity branches  $L\sub{max}=\floor{B/B\sub{c}}$, which is $4$ for the EPA 5\,Hz and $30$ for the TDL-C 300\,ns--3\,km/h. 
Throughout this section, we focus on Rayleigh fading, i.e., $\rmatH_{\ell}$ in~\eqref{eq:io} has \iid $\jpg(0,1)$ entries.
Since the noise variance is also $1$, we can interpret $\snr$ as the  SNR at each receive antenna.

%


In order to obtain an equivalent block-fading model, we limit the number of RBs per coherence bandwidth to $r\leq B\sub{c}/B\sub{RB}$.
The size of the coherence interval $\cohtime$ in~\eqref{eq:io} is thus $\cohtime=12dr$.
Choosing $L=4$ for the EPA 5\,Hz results for example in $n\sub{c}=72$, which can be obtained by setting $d=2$ and $r=3$. 
Similarly, choosing $L=12$ for the TDL-C 300\,ns--3\,km/h results in $n\sub{c}=24$, which corresponds to $d=2$ and $r=1$.
%
%

\begin{figure}\centering%
\includegraphics{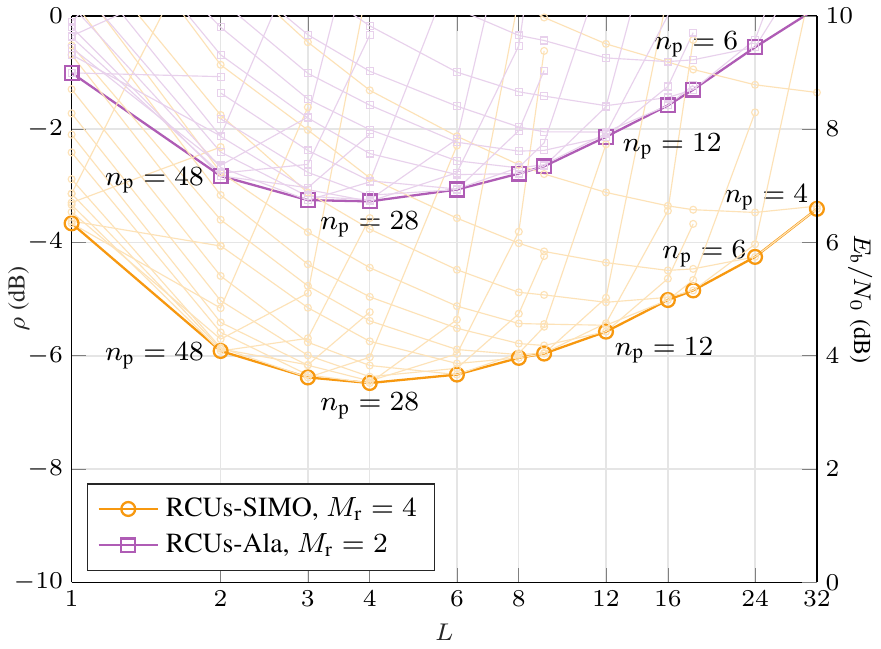}
\caption{Minimum SNR to achieve $\epsilon=10^{-3}$ as a function of the number of frequency diversity branches  used: each curve corresponds to a different number of pilot symbols. The optimal number of pilots is reported near some of the points of the envelope.}%
\label{fig:fig0}
\end{figure}

To illustrate how performance is affected by the choice of the number of diversity branches, we depict in Fig.~\ref{fig:fig0} the minimum SNR needed to achieve $10^{-3}$ as a function of the number of available diversity branches for the SIMO $1\times 4$ and the Alamouti $2\times 2$.
Each shaded curve in the plot corresponds to a different number of pilot symbols, and the envelope corresponds to the optimal number of pilots. 
We observe that $L=4$ yields the lowest SNR value for both systems.
However, the curves are rather flat around their minimum. 
For example, $\rho$ lies within $0.5\dB$ from its minimum value for all $L$ between $2$ and $9$.

%
\newlength{\figH}\newlength{\figW}%
\setlength\figH{0.65\columnwidth}\setlength\figW{0.90\columnwidth}%


\begin{figure}\centering%
\includegraphics{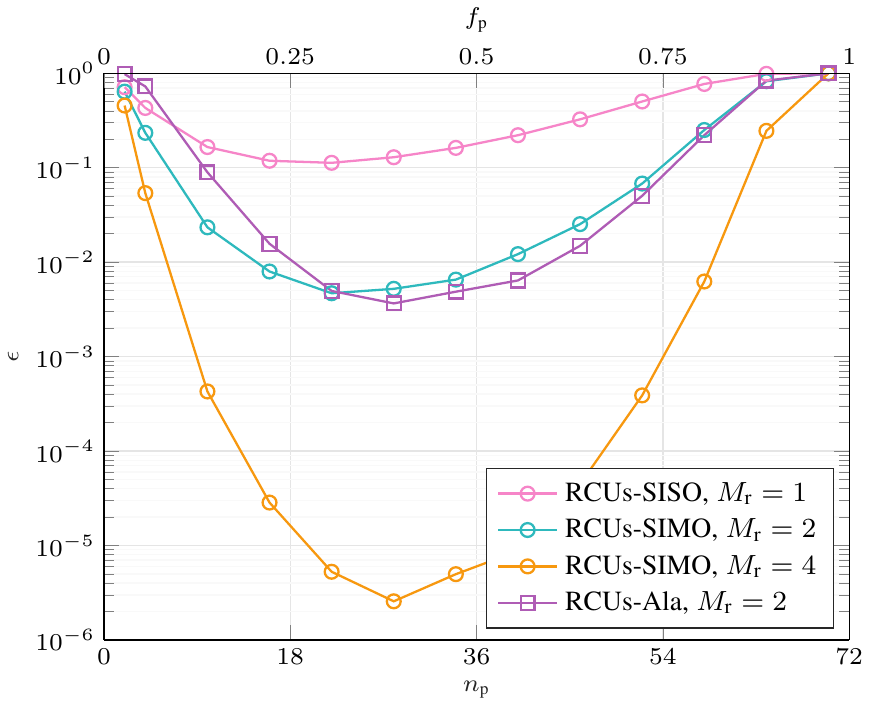}
\caption{Error probability vs. number of pilots (bottom axis) or fraction of pilots (top axis) for $\rho=-4$ dB (equivalent to $E\sub{b}/N_{0}=6$ dB) for the EPA 5\,Hz block-equivalent model.}%
\label{fig:fig1}
\end{figure}%
\begin{figure}\centering%
\includegraphics{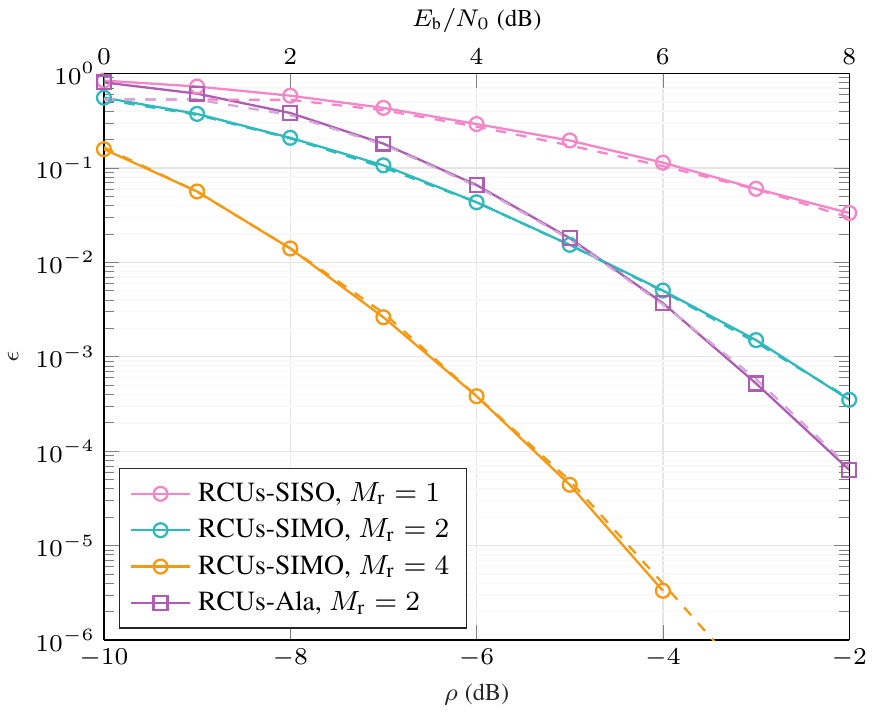}
\caption{Error probability vs. SNR (bottom axis) or $E\sub{b}/N_{0}$ (top axis) for the EPA 5\,Hz block-equivalent model. 
Solid lines: RCUs bound~\eqref{eq:RCUs_error}; dashed lines: saddlepoint approximation~\eqref{eq:SP_RCUs}.}
\label{fig:fig3}
\end{figure}%

We next focus on the EPA 5\,Hz channel and plot in Figs.~\ref{fig:fig1} and \ref{fig:fig3} the packet error probability as a function of the number of pilot symbols and the SNR, respectively.
Motivated by our findings in Fig.~\ref{fig:fig0}, we consider the case $L=4$, which is the maximum amount of frequency diversity offered by this channel (see Table~\ref{tab:channel-parameters}).
In Fig.~\ref{fig:fig1}, we report the error probability as a function of the number of pilot symbols for  SIMO  and  Alamouti.
Here, $\rho=-4\dB$. 
We observe that the error probability is extremely sensitive to changes in the number of pilot symbols. 
For example, in the SIMO $1\times 4$ case, reducing the number of pilot symbols from its optimal
 value of $28$ to $22$, which corresponds to a reduction in the fraction $f\sub{p}=\npilot/\cohtime$ of pilot symbols of $8.3\%$, doubles the error probability. 

In Fig.~\ref{fig:fig3}, we plot the error probability for the optimal fraction of pilot symbols found in Fig.~\ref{fig:fig1}. Each curve is computed for the optimal fraction of pilots of the corresponding scheme.
We also depict the corresponding saddlepoint approximations~\eqref{eq:SP_RCUs}, which turn out to be extremely accurate. 
As expected, the SIMO $1\times 4$ and the Alamouti $2\times 2$ curves have the same slope because these two setups provide the same amount of space-frequency diversity. 
The gap is around $3\dB$---the expected gap for the case of perfect CSIR. 
This implies that the channel estimate is sufficiently accurate.
We conclude from Figs.~\ref{fig:fig1} and \ref{fig:fig3} that the only system able to meet the target packet error probability of $10^{-5}$ within the range of SNR values considered in the figures is the SIMO $1\times 4$.
\begin{figure}\centering%
\includegraphics{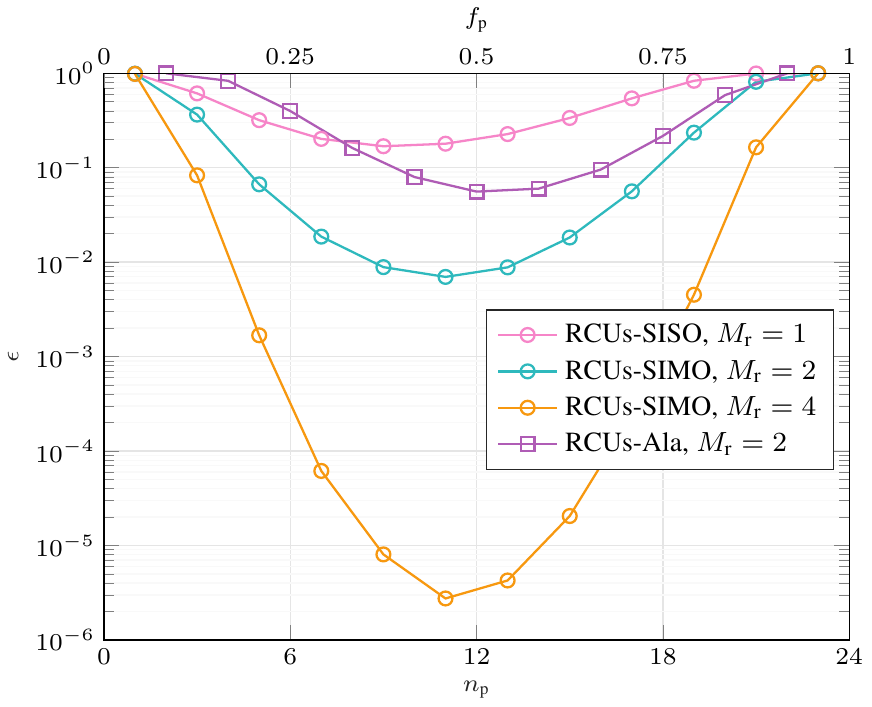}
\caption{Error probability vs. number of pilots (bottom axis) or fraction of pilots (top axis) for $\rho=-4$ dB (equivalent to $E\sub{b}/N_{0}=6$ dB for the TDL-C 300\,ns--3\,km/h block-equivalent model.}%
\label{fig:fig2}
\end{figure}%
%
%
\begin{figure}\centering%
\includegraphics{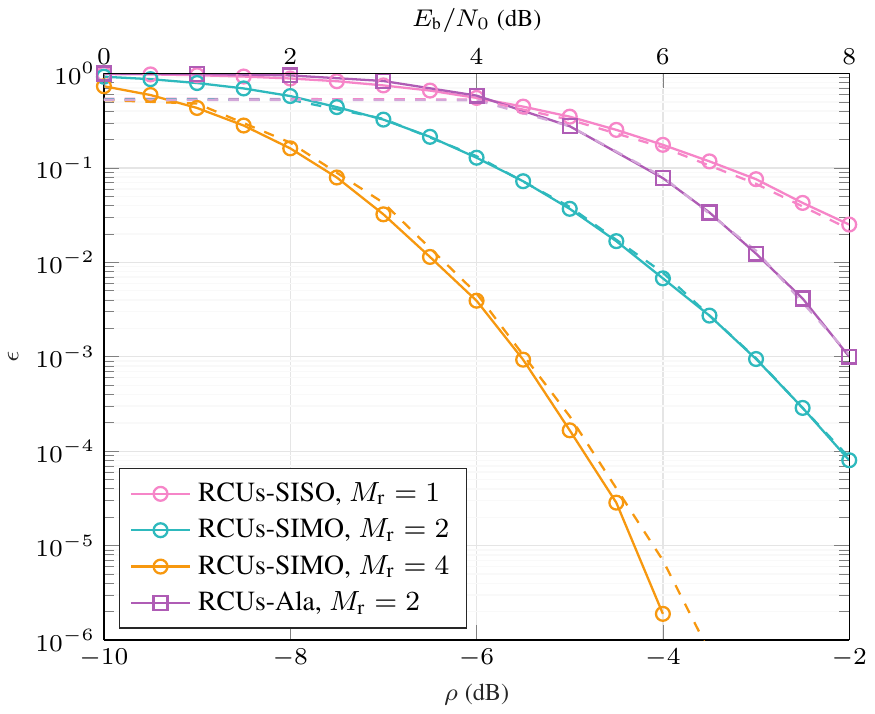}
\caption{Error probability vs. SNR (bottom axis) or $E\sub{b}/N_{0}$ (top axis) for the TDL-C 300\,ns--3\,km/h block-equivalent model.
Solid lines: RCUs bound~\eqref{eq:RCUs_error}; dashed lines: saddlepoint approximation~\eqref{eq:SP_RCUs}.}%
\label{fig:fig4}
\end{figure}%

We now move to the TDL-C 300\,ns--3\,km/h, which offers a larger maximum number of diversity branches in frequency. 
We assume that the system is designed so that $L=12$ (see Table~\ref{tab:channel-parameters} for a complete list of system parameters). 
Although the analysis reported in Fig.~\ref{fig:fig0} points out that $L=4$ should be chosen to minimize the SNR (at least at a  target packet error probability of $10^{-3}$), investigating the case $L=12$ allow us to assess the impact of frequency selectivity on the performance of the SIMO and the Alamouti schemes. 

Figs.~\ref{fig:fig2} and \ref{fig:fig4} parallel Figs.~\ref{fig:fig1} and \ref{fig:fig3}, respectively, with $L=12$ instead of $L=4$. 
Comparing Figs.~\ref{fig:fig1} and \ref{fig:fig2}, we see that, although the optimal number of pilot symbol is smaller when $L=12$, because the coherence block is smaller, the fraction of pilot symbols $f\sub{p}$ is actually larger.
We also observe that when the number of pilot symbols is chosen optimally, the minimum error probability achievable in the SIMO case when $L=12$  is similar  to when $L=4$.
On the contrary, the error probability of the Alamouti $2\times 2$ scheme increases by an order of magnitude when moving from $L=4$ to $L=12$.
This is because the Alamouti scheme is more sensitive to imperfect channel estimation, due to the processing needed to extract diversity from the two transmit antennas, i.e., the left-multiplication by  the matrices $\herm{\widehat{\rmatV}}_{\ell,1}$ and $\herm{\widehat{\rmatV}}_{\ell,2}$ in~\eqref{eq:equivalent_channel_alamouti}.
This effected is also illustrated in Fig.~\ref{fig:fig4}, where we see that the gap between  Alamouti and  SIMO $1\times 4$  is now $3.5\dB$, instead of the $3\dB$ loss we observed in Fig.~\ref{fig:fig3}.
Indeed, for this scenario better performance can be achieved within the range of SNR values depicted in the figure using a SIMO $1\times 2$, i.e., switching off one of the two transmit antennas.

\linespread{1} 
\bibliographystyle{IEEEtran}
\bibliography{IEEEabrv,publishers,confs-jrnls,refs,./Inputs/ref_giu}
\end{document}